\documentclass[Physsubmission, Phys]{SciPost}

\hypersetup{
    colorlinks,
    linkcolor={red!50!black},
    citecolor={blue!50!black},
    urlcolor={blue!80!black}
}

\usepackage[bitstream-charter]{mathdesign}
\DeclareSymbolFont{usualmathcal}{OMS}{cmsy}{m}{n}
\DeclareSymbolFontAlphabet{\mathcal}{usualmathcal}

\begin{document}

\begin{center}{\Large \textbf{
Determination of the rapidity evolution kernel from\\
  Drell-Yan data at low transverse momenta  
}}\end{center}

% TODO: write the author list here. Use initials + surname format.
% Separate subsequent authors by a comma, omit comma at the end of the list.
% Mark the corresponding author with a superscript *.
\begin{center}
F. Hautmann\textsuperscript{1, 2},
I. Scimemi\textsuperscript{3} and
A. Vladimirov\textsuperscript{4}
\end{center}

\begin{center}
{\bf 1} Elementaire Deeltjes Fysica, Universiteit Antwerpen, B 2020 Antwerpen
\\
{\bf 2} Theoretical Physics Department, University of Oxford, Oxford OX1 3PU
\\
{\bf 3} Departamento de Física Teórica and IPARCOS, Facultad de Ciencias Físicas, Universidad Complutense Madrid, Plaza Ciencias 1, 28040 Madrid, Spain
\\
{\bf 4} Institut f\"ur Theoretische Physik, Universit\"at Regensburg, D-93040 Regensburg, Germany
%* correspondingauthor@email.address 
\end{center}

%\begin{center}
%\today
%\end{center}

%\linenumbers

\definecolor{palegray}{gray}{0.95}
\begin{center}
\colorbox{palegray}{
  \begin{tabular}{rr}
  \begin{minipage}{0.1\textwidth}
    \includegraphics[width=22mm]{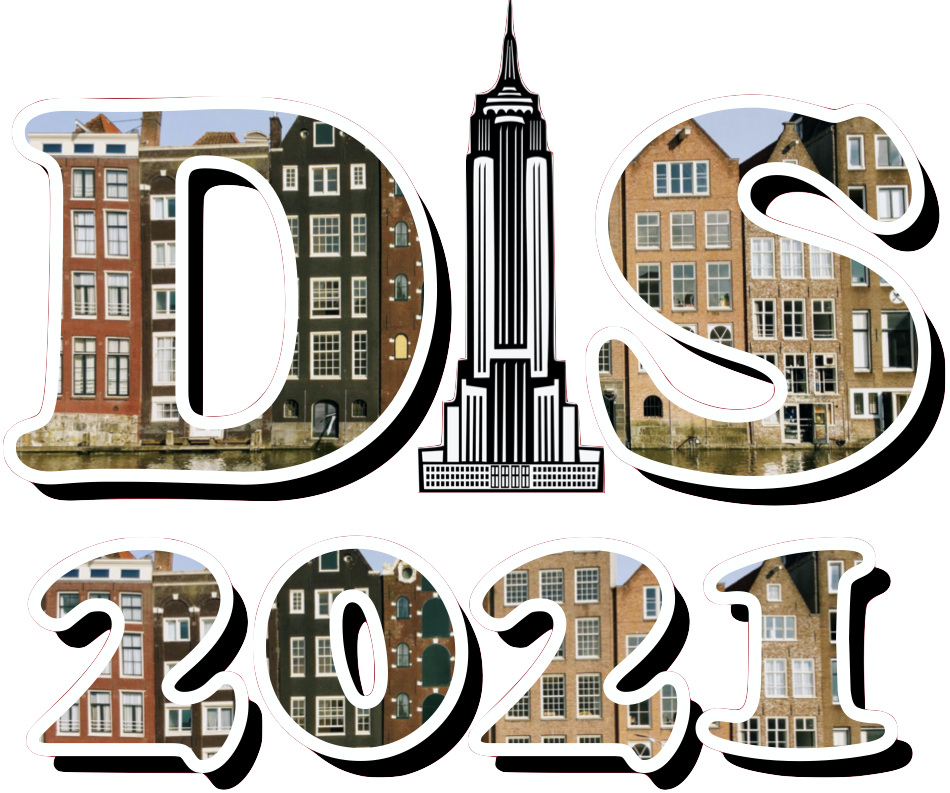}
  \end{minipage}
  &
  \begin{minipage}{0.75\textwidth}
    \begin{center}
    {\it Proceedings for the XXVIII International Workshop\\ on Deep-Inelastic Scattering and
Related Subjects,}\\
    {\it Stony Brook University, New York, USA, 12-16 April 2021} \\
    \doi{10.21468/SciPostPhysProc.?}\\
    \end{center}
  \end{minipage}
\end{tabular}
}
\end{center}

\noindent\fbox{%
    \parbox{\textwidth}{%
\section*{Abstract}
{\bf   
We report  recent results on the extraction of the rapidity evolution kernel for Sudakov processes from 
experimental measurements of  Drell-Yan (DY) transverse momentum distributions. We 
discuss  the role of  high-precision   Large Hadron Collider (LHC) data,  and illustrate 
 the interplay of this analysis  with  transverse momentum dependent and collinear parton density functions. 
}
}%
}

\vspace*{1.6cm}

The experimental program of high-precision measurements   in DY lepton-pair production at the LHC 
influences   determinations of  the rapidity evolution kernel for Sudakov processes in QCD,  complementing    
   the information which may be obtained  from lower-energy DY experiments as well as semi-inclusive deep inelastic scattering.  
The kernel can be extracted from  precise measurements of the DY transverse momentum distribution in the region of low 
transverse momenta.  
In this article we  discuss recent results on this  based on our work~\cite{Hautmann:2020cyp}.

The rapidity evolution kernel was introduced by Collins and Soper \cite{Collins:1981uk,Collins:1981va} in the 
context of transverse momentum dependent (TMD) 
factorization \cite{Collins:1984kg} for the DY differential cross section at low transverse 
momenta \cite{Dokshitzer:1978hw,Parisi:1979se,Curci:1979bg}.  It controls the evolution in rapidity of 
partonic TMD  distribution functions  \cite{Angeles-Martinez:2015sea}  as a function of mass $\mu$ and 
transverse coordinate  $b_T$.  
Rapidity divergences  are  treated by cut-off methods  \cite{Collins:1981uk,Collins:1984kg} 
or by  subtraction methods \cite{Collins:1999dz,Collins:2003fm,Hautmann:2007uw,Collins:2011zzd}. 
The kernel's rate of change with mass is given 
by the perturbatively-calculable cusp anomalous dimension (currently known numerically to four loops \cite{Moch:2018wjh,Moch:2017uml}). 
For large transverse distances 
$b_T  {\raisebox{-.6ex}{\rlap{$\,\sim\,$}} \raisebox{.4ex}{$\,>\,$}} \Lambda^{-1}_{\rm{QCD}}$ 
the kernel is non-perturbative. 

Information on the non-perturbative rapidity evolution kernel may be obtained by lattice QCD methods. Studies 
in this direction have recently 
started \cite{Ebert:2019tvc,Ebert:2018gzl,Schlemmer:2021aij,LatticeParton:2020uhz,Shanahan:2021tst,Shanahan:2020zxr}.   
Alternatively, it may be obtained by comparison 
of theory with experimental measurements. This has a long history, from the pioneering 
`ResBos' fits \cite{Ladinsky:1993zn,Landry:1999an,Landry:2002ix,Konychev:2005iy} of Tevatron data to the 
recent  `artemide' \cite{Scimemi:2019cmh} and `NangaParbat' \cite{Bacchetta:2019sam} global fits to 
 LHC, Tevatron, RHIC  and fixed-target data. 

The large-distance  behavior of the rapidity evolution kernel, 
currently  subject of  active investigation,   
  has traditionally been  modeled as a quadratic 
behavior $\propto b_T^2$.  The fits \cite{Ladinsky:1993zn,Landry:1999an,Landry:2002ix,Konychev:2005iy}, for instance, 
depend on such model. The analysis \cite{Hautmann:2020cyp}, emphasizing the bias due to this assumption, examines 
alternative scenarios.  One of these scenarios has the perturbative quadratic behavior $\propto b_T^2$ at small  $b_T$ 
but a constant behavior $\propto b_T^0$ at large $b_T$, fulfilling the asymptotic condition $\partial {\cal D}  / \partial 
\ln b_T^2  = 0$  for the kernel ${\cal D}$, 
in a similar spirit to parton saturation in the $s$-channel picture \cite {Hautmann:2007cx,Hautmann:2000pw} for 
partonic distribution functions.   Another scenario, intermediate between the `traditional' and the `saturating' ones, 
has a linear behavior $\propto b_T^1$ at large $b_T$.  We discuss below the results  of 
fits based on  these three different scenarios with, respectively,  quadratic, linear or constant asymptotic large-$b_T$ behaviors.

The determination of the rapidity evolution kernel is intertwined with the determination of the non-perturbative 
TMD parton distributions. While the 
evolution in mass of TMD distributions is controlled by   perturbatively-calculable 
 kernels,   the TMD distributions at a fixed (low) mass scale are non-perturbative and may be 
 determined from fits to experiment. This is analogous to the case of ordinary, collinear parton density functions (PDF) and for this 
 reason, similarly to the PDF case, libraries of TMDs such as \cite{Abdulov:2021ivr,Hautmann:2014kza}, obtained from 
 parameterizations and  fits to data,   are a necessary tool for phenomenology.  The non-perturbative TMDs at low mass scales provide 
 information about the `intrinsic transverse momentum $k_T$' in the hadron and  `3-dimensional' (3D) imaging of hadron structure.   
  We  discuss below the results 
 of simultaneous fits of the non-perturbative rapidity evolution kernel and TMD distributions. In this respect, it is also interesting to 
 ask  how the determination of the rapidity evolution kernel is affected  if the 3D structure is disregarded, that is, if the non-perturbative 
  TMD distribution is simply taken to be a $\delta$-function in $k_T$. Thus, besides the simultaneous fits, we  also consider the 
  alternative scenario of fitting the   kernel while assuming $\delta$-function TMD distributions.

\begin{figure}[t]
\begin{center}
\includegraphics[width =0.4\textwidth]{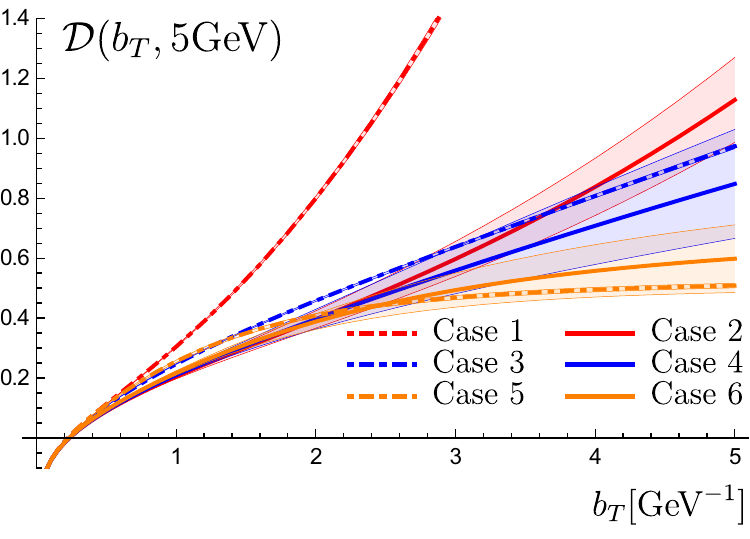}
\hspace{2cm}
\includegraphics[width =0.4\textwidth]{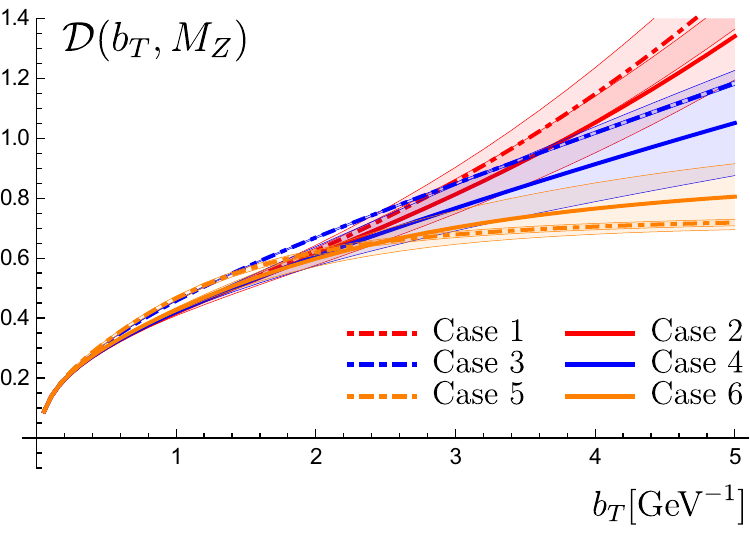}
\end{center}
\caption{ \textcolor{black}{
Rapidity  evolution kernel \cite{Hautmann:2020cyp} 
as a function of $b_T$ for  different values of mass,   
 $\mu=5$ GeV (left) and  $\mu=M_Z$ (right), obtained from fits to LHC DY experimental data.  
 The different curves  correspond to the scenarios described in the text (see \cite{Hautmann:2020cyp} for further details).  
}
\label{fig:D}
}
\end{figure}
 
Fig.~\ref{fig:D} presents the results of the analysis \cite{Hautmann:2020cyp} for the rapidity evolution kernel   ${\cal D}$. 
 The analysis is based on  next-to-next-to-leading-logarithmic (NNLL) order predictions from the 
 `artemide' code \cite{Hautmann:2020cyp,Scimemi:2019cmh}, fitted to experimental 
 data for the DY transverse momentum $q_T$ distribution from the 
 LHC~\cite{Aad:2014xaa,Aad:2015auj,Chatrchyan:2011wt,CMS:2016mwa,Aaij:2015gna,Aaij:2015zlq,Aaij:2016mgv} 
in the low transverse momentum region specified by the cut $q_T / Q \leq \kappa $, where $Q$ is the DY invariant mass and 
$\kappa$ is taken to be 0.2.\footnote{We have varied the cut $\kappa$ on the data set in the range from 0.1 to 0.25 and 
verified that the results for the non-perturbative ${\cal D}$ and TMD parameters are stable in this range.  The calculation 
cannot be used for  $\kappa$ near  1  as it does not 
include  the matching of TMD-factorized contributions with next-to-leading  and higher order contributions to hard scattering matrix 
elements. See e.g. \cite{Bacchetta:2019tcu,BermudezMartinez:2019anj,BermudezMartinez:2020tys}   for recent discussions of DY measurements  
with $ q_T \sim Q$ in the context of analytic \cite{Collins:1984kg} and Monte Carlo \cite{Collins:2000gd}  matching.}  
  The odd-numbered curves (dashed lines) in Fig.~\ref{fig:D}  correspond to fits performed in 
the three scenarios for the large-$b_T$ asymptotics described above (1: quadratic; 3: linear; 5: constant) with $\delta$-function TMDs, 
while the even-numbered curves (solid lines) correspond to simultaneous  ${\cal D}$ and TMD fits in the three asymptotics 
scenarios (2: quadratic; 4: linear; 6: constant).

We see from Fig.~\ref{fig:D} that while the kernel ${\cal D}$ is rather well determined for $ b_T  
{\raisebox{-.6ex}{\rlap{$\,\sim\,$}} \raisebox{.4ex}{$\,<\,$}} $ 2   GeV$^{-1}$ using LHC DY $q_T$ data,  
the sensitivity of current LHC measurements to the long-distance region of higher  $ b_T $ is limited,  
 which results into sizeable uncertainty bands for $ b_T 
{\raisebox{-.6ex}{\rlap{$\,\sim\,$}} \raisebox{.4ex}{$\,>\,$}} $ 2 - 3 GeV$^{-1}$. It is shown in \cite{Hautmann:2020cyp} that the  
scenarios with simultaneous ${\cal D}$ and TMD fits  (even-numbered  cases in Fig.~\ref{fig:D}) 
 all give good $\chi^2$, with comparable $\chi^2$ values,  while  the  corresponding results for 
  ${\cal D}$ in Fig.~\ref{fig:D} differ significantly.  
  
We also see  from Fig.~\ref{fig:D} that the correlation between the kernel ${\cal D}$ and the non-perturbative TMD is significant at 
large $ b_T $. This is measured by the difference between each of the even-numbered curves and its odd-numbered counterpart 
which has the same large-$b_T$ asymptotic behavior but starts with a $\delta$-distribution in $k_T$ before TMD evolution sets in. 
The difference is especially large for the quadratic case in Fig.~\ref{fig:D} (left) for $\mu=5$ GeV, even exceeding the uncertainty bands. 
It is observed in \cite{Hautmann:2020cyp} that it is possible to obtain good fits to LHC DY data with  some of the 
$\delta$-distribution scenarios, but  inducing significant biases  in the   ${\cal D}$ determination.

\begin{figure}[h]
\begin{center}
\includegraphics[width =0.25\textwidth]{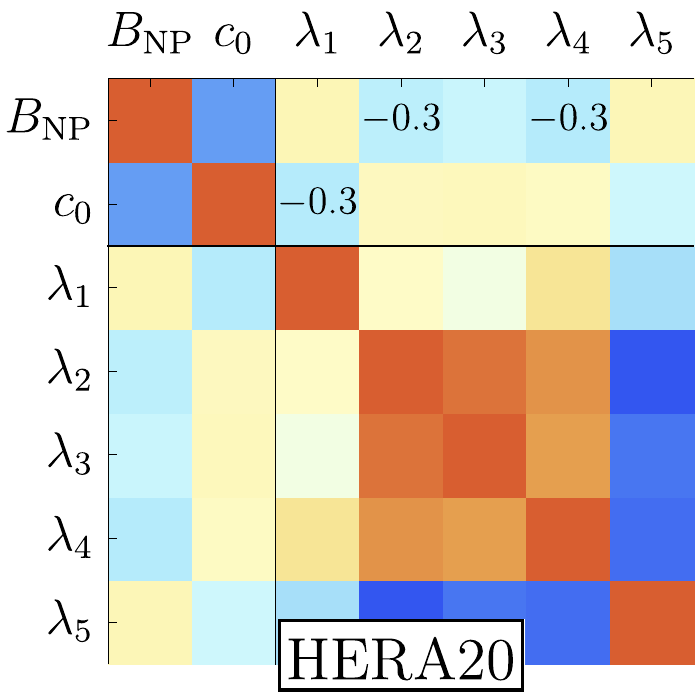}
\hspace{0.5cm}
\includegraphics[width =0.25\textwidth]{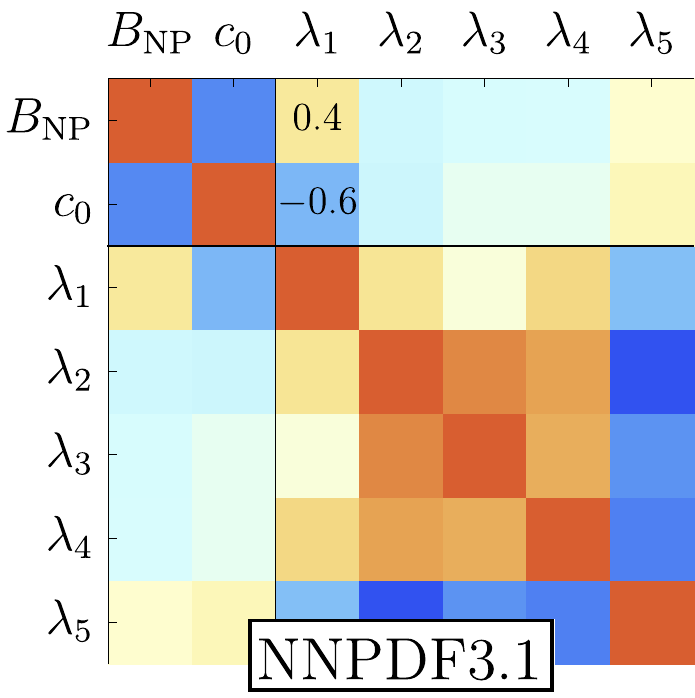}
\hspace{0.5cm}
\includegraphics[width =0.04\textwidth]{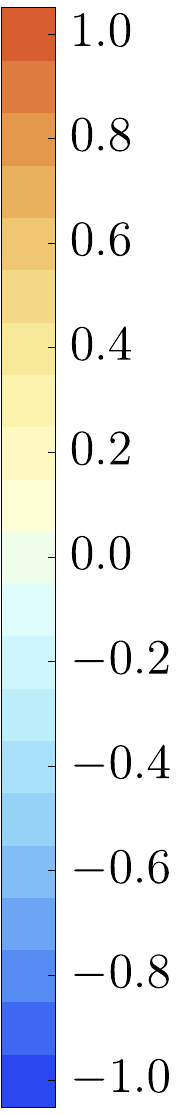}
\\
\includegraphics[width =0.25\textwidth]{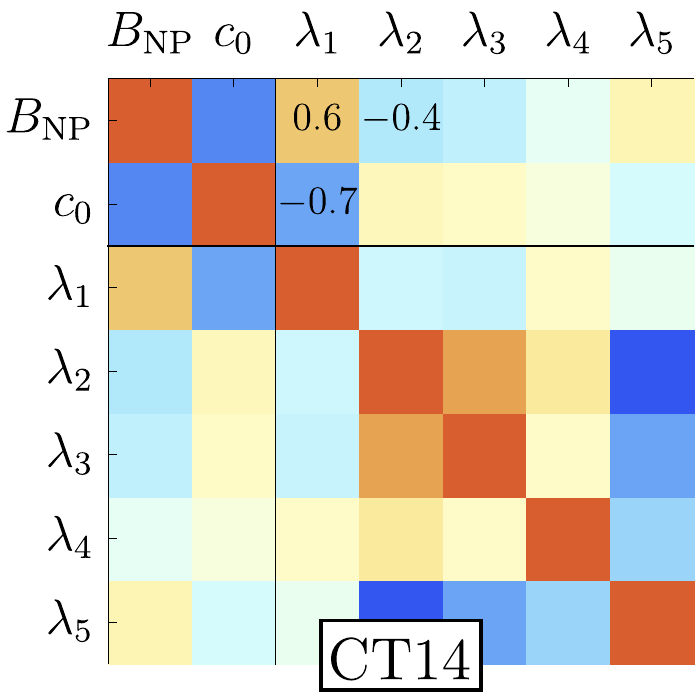}
\hspace{0.5cm}
\includegraphics[width =0.25\textwidth]{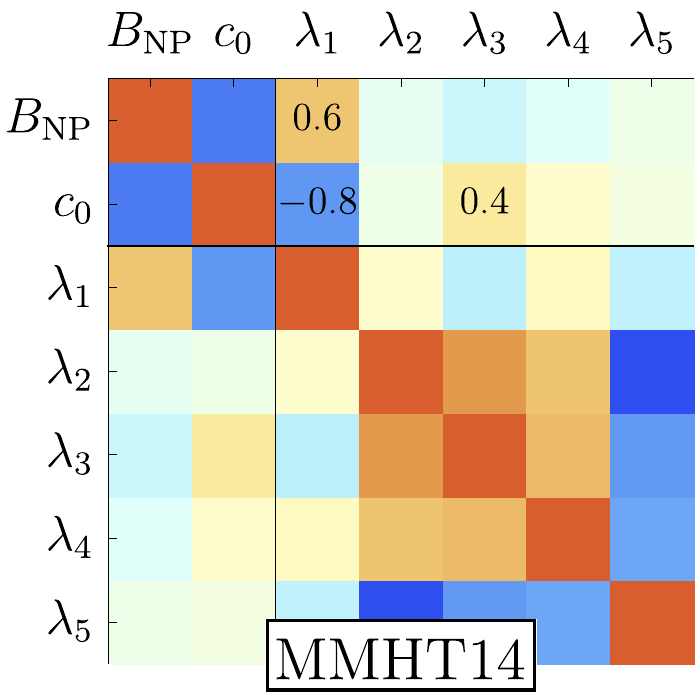}
\hspace{0.5cm}
\includegraphics[width =0.25\textwidth]{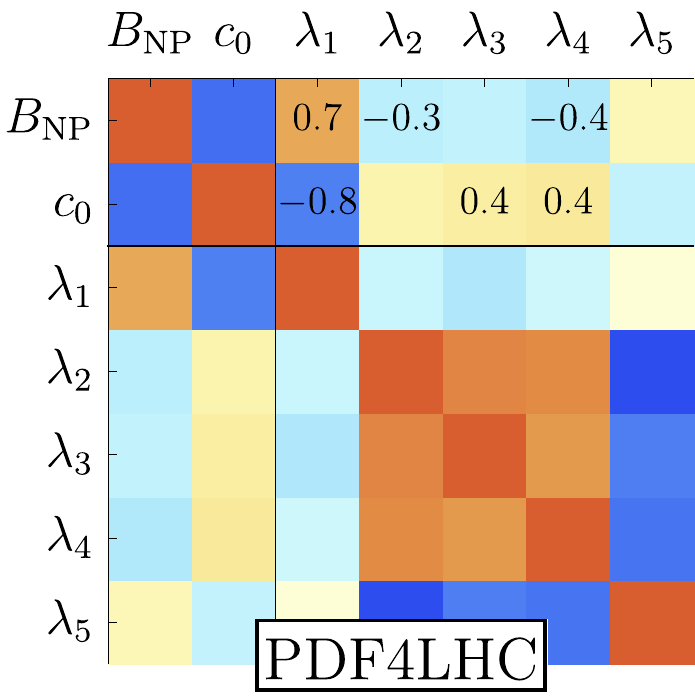}
\hspace{0.5cm}
\end{center}
\caption{Correlations of ${\cal D}$ ($B_{NP}, c_0 $) and TMD ($\lambda_i$) 
 fit parameters \cite{Hautmann:2020cyp}  for different PDF sets. 
 In the axes $1=B_{NP}$, $2=c_0$, $(3,4,5,6,7)=\lambda_{1,2,3,4,5}$. 
Light colors indicate low correlations; dark colors  indicate 
high correlations. Shades of blue 
denote negative correlations; shades of brown denote positive correlations.  
 (Diagonal entries are trivial.) 
\label{fig:correlations}}
\end{figure}

To improve the uncertainties seen at large $b_T$, one possibility is to use data from lower-energy DY 
experiments \cite{PHENIX:2018dwt,Aaltonen:2012fi,Affolder:1999jh,Abazov:2010kn,Abazov:2007ac,Abbott:1999wk,McGaughey:1994dx,Moreno:1990sf,Ito:1980ev}. 
The sensitivity to large $b_T$ here increases compared to LHC data, but the precision decreases. 
In \cite{Hautmann:2020cyp} a global TMD fit is presented to DY data from LHC and lower-energy experiments. 
Useful constraints also come from low-energy semi-inclusive deep inelastic scattering (SIDIS) measurements \cite{Scimemi:2019cmh}.\footnote{This 
is however at the price  of additional non-perturbative parameters from TMD fragmentation. See e.g. discussions in 
  \cite{Proceedings:2020eah,LHeC:2020van} in the context of future lepton-hadron collider programs.}  
A determination of the rapidity evolution kernel from  combined LHC + low-energy data, obtained by  ${\cal D}$ and TMD fits 
and assuming linear large-$b_T$ behavior, is found \cite{Hautmann:2020cyp}   to be 
lower than its LHC-only analogue, curve 4 in Fig.~\ref{fig:D}, and close to curve 6 within uncertainties. 
Another possibility is for future LHC experiments to extend the kinematic 
reach to regions of low DY masses in which the  non-perturbative sensitivity   increases, that is,   
measure   the DY  $q_T$, for low $q_T \ll Q$ 
and  with fine binning $ 
{\raisebox{-.6ex}{\rlap{$\,\sim\,$}} \raisebox{.4ex}{$\,<\,$}}   1 $~GeV,  in   the so far 
unexplored region  $ Q <  40 $ GeV by lowering lepton $p_T$ thresholds.    

At the same time, improving our knowledge  of the rapidity evolution kernel will require  that 
 theoretical systematic uncertainties be under control. On one hand, these include perturbative uncertainties 
 related  to the expansions in the coupling $\alpha_s$ and the all-order logarithmic resummation. These can be evaluated by 
 examining the dependence of the results on  resummation scales as well as renormalization and factorization scales. 
On the other hand,  uncertainties from PDFs, i.e., nonperturbative contributions of collinear origin,   constitute a significant 
source of theoretical systematics as well. 
Fig.~\ref{fig:correlations}   illustrates 
the correlations among TMD parameters in the analysis \cite{Hautmann:2020cyp} 
for different  PDF sets  \cite{Ball:2017nwa,Abramowicz:2015mha,Dulat:2015mca,Harland-Lang:2015nxa,Butterworth:2015oua}. 
We see for instance  that  the correlation between the parameters 
$c_0$ (controlling the long-distance behavior of the rapidity evolution kernel) and $\lambda_1$ (controlling the intrinsic transverse 
momentum distribution) is  fairly low 
in the case of the HERAPDF set, but it increases in the NNPDF3.1 case, and is 
higher still in the CT14 and MMHT14 cases. 
Therefore, taking into account  the role of collinear 
PDFs and their uncertainties systematically in TMD fits is  one of the essential inputs  to  advances in the determination of  
 the rapidity evolution kernel for Sudakov processes. 
 Acknowledgements.
This study was supported by Deutsche Forschungsgemeinschaft (DFG) through the research Unit FOR 2926, “Next Generation pQCD for Hadron Structure: Preparing for the EIC”, project number 30824754. I.S. is supported by the Spanish Ministry grant PID2019-106080GB-C21. 
This project has received funding from the European Union Horizon 2020 research and innovation program under grant agreement Num. 824093 (STRONG-2020).

\bibliography{nonpDISbib24sept}

\nolinenumbers

\end{document}